\documentclass[%
superscriptaddress,
showkeys,
 amsmath,amssymb,
 aps,prfluids
]{revtex4-2}

\usepackage{graphicx}
\usepackage{dcolumn}
\usepackage{bm}
\usepackage{xcolor}
\usepackage{amsmath}

\usepackage[normalem]{ulem}
\usepackage{cancel}

\begin{document}
\preprint{APS/123-QED}

\title{Short- and long-term prediction of a chaotic flow: A physics-constrained reservoir computing approach}

\author{N.A.K. Doan}
\affiliation{%
 Department of Mechanical Engineering, Technical University of Munich, Garching 85747, Germany}%
\affiliation{%
 Institute for Advanced Study, Technical University of Munich, Garching 85748, Germany}%
\author{W. Polifke}%
\affiliation{%
 Department of Mechanical Engineering, Technical University of Munich, Garching 85747, Germany}%

\author{L. Magri}\email{lm547@cam.ac.uk}
\affiliation{
 Department of Engineering, University of Cambridge, Cambridge CB2 1PZ, United Kingdom
}%
\affiliation{%
 Institute for Advanced Study, Technical University of Munich, Garching 85748, Germany (visiting)}%
\affiliation{%
The Alan Turing Institute, 96 Euston Road, London, England, NW1 2DB, United Kingdom}%
 
\date{\today}

\begin{abstract}
We propose a physics-constrained machine learning method---based on reservoir computing--- to time-accurately predict extreme events and long-term velocity statistics in a model of turbulent shear flow. The method leverages the strengths of two different approaches: empirical modelling based on reservoir computing, which {\it learns} the chaotic dynamics from data only, and physical modelling based on conservation laws, which {\it extrapolates} the dynamics when training data becomes unavailable. We show that the combination of the two approaches is able to accurately reproduce the velocity statistics and to predict the occurrence and amplitude of extreme events in a model of self-sustaining process in turbulence. In this flow, the extreme events are abrupt transitions from turbulent to quasi-laminar states, which are deterministic phenomena that cannot be traditionally predicted because of chaos.  Furthermore, the physics-constrained machine learning method is shown to be robust with respect to noise.
This work opens up new possibilities for synergistically enhancing  data-driven methods with physical knowledge for the {time-accurate} prediction of  chaotic flows. 
\end{abstract}

\maketitle

\section{Introduction}
Many fluid dynamics systems exhibit extreme events, which are violent and sudden changes of a flow from the average evolution \citep{Farazmand2019}. 
Examples of extreme events in fluids are oceanic rogue waves~\citep{Dysthe2008}, extreme patterns in atmospheric and climate science \citep{Easterling2000,Majda2012}, intermittency in turbulence~\citep{Platt1991}, and thermoacoustic instabilities in aeroengines and rocket motors~\citep{Lieuwen2006a}, to name only a few. 
Although governed by deterministic equations such as conservation laws, extreme events occur in a seemingly random way. 
The time-accurate detection and prediction of the dynamics of extreme events can be achieved for a  short time scale, which is the predictability time~\citep{Boffetta2002}.  
This is a roadblock for the {\it time-accurate} prediction because, after the predictability time, a tiny difference between the initial conditions, such as floating-point errors, is exponentially amplified. 
This is  popularly known as the {butterfly effect} in chaos theory~\citep{Lorenz1963}. 
Because of this, the time-accurate prediction of extreme events remains an open problem~\citep{Farazmand2019}. 

Most state-of-the-art predictive approaches rely on statistical methods. Extreme Value Theory~\citep{Nicodemi2015}  and Large Deviation Theory~\citep{Varadhan2008} characterize the probability of an event and the heavy tail of the probability density function, which can  be used to compute the initial conditions with the highest probability of transitioning towards extreme values. 
Statistical approaches successfully identified precursors of turbulent channel flow relaminarization~\citep{Blonigan2019} and of nonlinear rogue waves~\citep{Dematteis2018}. 
Another approach for the prediction of extreme events is data-driven. Machine learning has proved successful at predicting the dynamics of some chaotic flows with accurate short-term predictions and long-term statistics reproduction. \citet{Vlachas2018} used a Long-Short Term Memory (LSTM) network, which is a type of Recurrent Neural Network (RNN) \citep{Goodfellow2016}, to predict the evolution of the Kuramoto-Sivashinsky equation and of a barotropic climate model. 
They showed that the proposed neural network had a good short-term accuracy, and converged towards the correct invariant measures.
A similar architecture was employed by \cite{Srinivasan2019} to simulate the evolution of a model of shear turbulence showing that it was capable of reproducing the moments of the velocity statistics. Another type of RNN---the Echo State Network \citep{Jaeger2007,Lukosevicius2009}---based on reservoir computing, can successfully learn the chaotic dynamics beyond the predictability time for short-term predictions~\citep{Pathak2018,Pathak2018a,Doan2019} and ergodic averages~\citep{Huhn2020}. 
Data-driven methods have the capability of predicting chaotic dynamics, but it is still an open question whether they can be robustly used to predict extreme events. 
\citet{Wan2018} tackled this problem with a hybrid approach that combined a reduced-order model with a LSTM. They predicted the evolution of dissipation events in the Kolmogorov flow and intermittent transitions between two flow regimes in a model of barotropic flow. \citet{Sapsis2018} combined Large Deviation Theory with a data-driven method to efficiently characterize the heavy tail of the distribution in the Kolmogorov flow. 

Statistical methods provide a robust framework to identify precursors and calculate the probability of extreme events, but they do not provide a robust way to time-accurately predict their occurrence and amplitude. On the other hand, recurrent neural networks, such as LSTMs and ESNs, can predict the dynamics of chaotic systems by learning temporal patterns in data only, but, because they are fully data driven, they provide solutions that may violate the governing physical laws, such as momentum conservation. In fluid mechanics, machine-learned solutions, however, should obey physical principles such as momentum and mass conservation. This calls for embedding physical and domain knowledge in data-driven methods~\citep[e.g., ][]{Karpatne2017a,Raissi2019b,Doan2019,Baker2019, Brunton2020}. 
The objective of this paper is to propose a machine learning method that (i) produces physical solutions to time-accurately predict extreme events in a qualitative model of shear turbulence, and (ii) reproduces the long-term statistics. 
The flow configuration is presented in Sec. \ref{sec:TurbModel}. The PI-ESN framework is developed in Sec. \ref{sec:PI-ESN}. Results are shown in Sec. \ref{sec:results}. The long-term statistical behaviour of the network is discussed in Sec. \ref{sec:res_LT}. Short-term predictions of extreme events are analysed in Sec. \ref{sec:res_EE}. The robustness of the overall architecture with respect to noise is presented in Sec. \ref{sec:res_noise}. A final discussion with future directions concludes the paper. 

 \section{Extreme events in a model of shear turbulence}
\label{sec:TurbModel}
 We regard the chaotic flow as an autonomous dynamical system
\begin{equation}
 \dot{\bm{y}} = \mathcal{N} (\bm{y}) \;\;\;\;\;\;\;\;\; \bm{y}(0)  =\bm{y}_0
\end{equation}
where 
$\dot{(\;)}$ is the temporal derivative; and $\mathcal{N}$ is a deterministic nonlinear  differential operator, which encapsulates the numerical discretization of the spatial derivatives and boundary conditions (if any). 
 The flow is governed by momentum and mass conservation laws, i.e., the Navier-Stokes equations, which were reduced in form by \cite{Moehlis2004} (supplementary material). This model, which was inspired by earlier works by \cite{Waleffe1995,Waleffe1997},  provides the nonlinear operator $\mathcal{N}$.  This is called the MFE model for brevity. The MFE model  captures the qualitative features of the transition from turbulence to quasi-laminar states such as the exponential distribution of turbulent lifetimes. The velocity field in the model is decomposed as
\begin{equation}
\bm{v}({\bm x},t) = \sum_{i=1}^9 a_i (t) \bm{v_i}(\bm{x}),
\end{equation}
where $\bm{v_i}(\bm{x})$ are spatial Fourier modes (or combinations of them)~\citep{Moehlis2004}. The Navier-Stokes equations are projected onto $\bm{v_i}(\bm{x})$ to yield nine ordinary differential equations for the modes' amplitudes, $a_i$, which are nonlinearly coupled. Consequently, the state vector is ${\bm y}=\{a_i\}_1^9$. All the variables are non-dimensional~\citep{Moehlis2004}.  
Physically, 
$\bm{v}_1$ is the laminar profile mode; 
$\bm{v}_2$ is the streak mode; 
$\bm{v}_3$ is the downstream vortex mode; 
$\bm{v}_4$ and $\bm{v}_5$ are the spanwise flow modes; 
$\bm{v}_6$ and $\bm{v}_7$ are the normal vortex modes; 
$\bm{v}_8$ is the three-dimensional mode; and 
$\bm{v}_9$ is the modification of the mean profile caused by turbulence.
The flow has a fixed point $a_1=1$, $a_2=...=a_9=0$, which is a laminar state. The supplementary material reports the expressions for $\bm{v_i}$ and the equations for $a_i$~\citep{Moehlis2004}.
Detailed analysis of the MFE model has been performed by \citet{Kim2008}  and \citet{Joglekar2015}. 

\begin{figure}[!ht]
    \centering
    \includegraphics[width=0.6\textwidth]{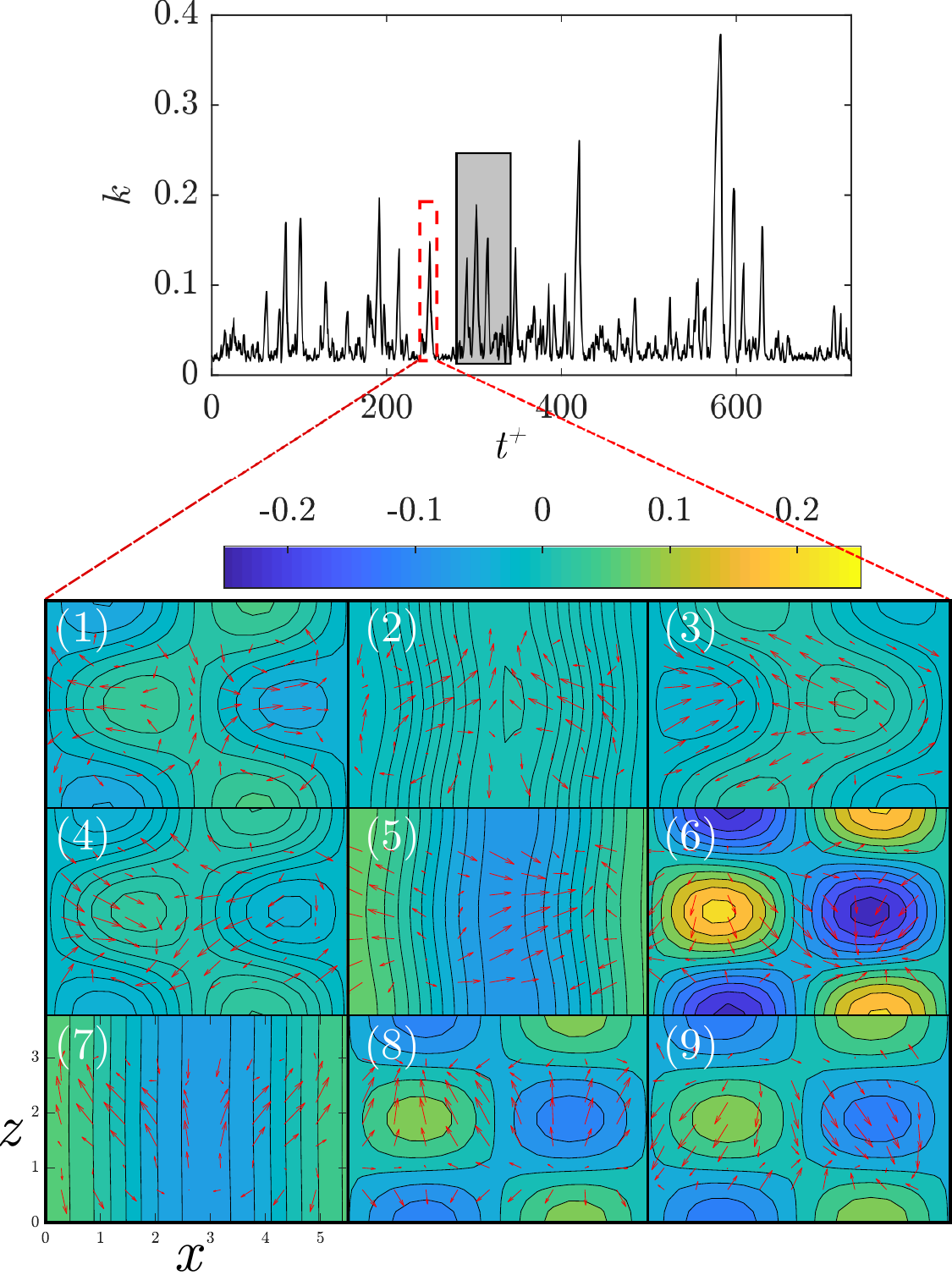}
    \caption{Top panel: kinetic energy, $k$. The grey box indicates the training window of the physics-informed echo state network (PI-ESN). Bottom panel: velocity field in the mid-$y$ plane. The arrows indicate the in-plane velocity ($x$-$z$ directions), and the colour maps indicate the ouf-of-plane velocity. $t^+$ is the time normalised by the largest Lyapunov exponent.%
    }
    \label{fig:MFE_evol}
\end{figure}

The flow under investigation is incompressible. The domain is a cuboid of size $L_x \times L_y \times L_z$ between two infinite parallel walls at $y=0$ and $y=L_y$, which are periodic in the $x$ and $z$ directions. The domain size is $L_x=1.75\pi$, $L_y=2$ and $L_z=1.2\pi$. The Reynolds number is $600$. A sinusoidal volume force is applied in the $y$-direction. The initial condition (supplementary material) is such that the turbulent flow displays chaotic bursts between the fully turbulent and quasi-laminar states. These are the extreme turbulent events we wish to predict.
The governing equations are integrated in time with a 4$^{th}$-order Runge-Kutta scheme with a time step $\Delta t = 0.25$ to provide the evolution of the nine modes, $a_i$, from $t=0$ to  $t=30000$.
The evolution of the kinetic energy, $k= 0.5 \sum_{i=1}^9 a_i^2$, is shown in Fig.~\ref{fig:MFE_evol}. The time is normalized by the largest Lyapunov exponent, $\lambda_{\max}=41$, which is calculated as the average logarithmic error growth rate between two nearby trajectories~\citep{Boffetta2002}. The Lyapunov time scale, $\lambda_{\max}^{-1}$, provides an estimate of the predictability time, which is used to define the nondimensional time
\begin{equation}
    t^+=\frac{t}{\lambda^{-1}_{\max}}
\end{equation}
The kinetic energy, $k$, has sudden large bursts that arise from a chaotic oscillation with a small amplitude. 
Each burst is a quasi-relaminarization event, which occurs in three phases (Fig.~\ref{fig:MFE_evol}):  
(i) the originally laminar velocity profile becomes unstable and breaks down into vortices due to the shear imposed by the volume force (panels 5-7);
(ii) the vortices align to become streaks (panels 8-9 and 1-2); and
(iii) the streaks break down leading to flow relaminarization (panels 3-5). 

\section{Physics-constrained reservoir computing}
\label{sec:PI-ESN}

\begin{figure}[!ht]
    \centering
    \includegraphics[width=0.6\textwidth]{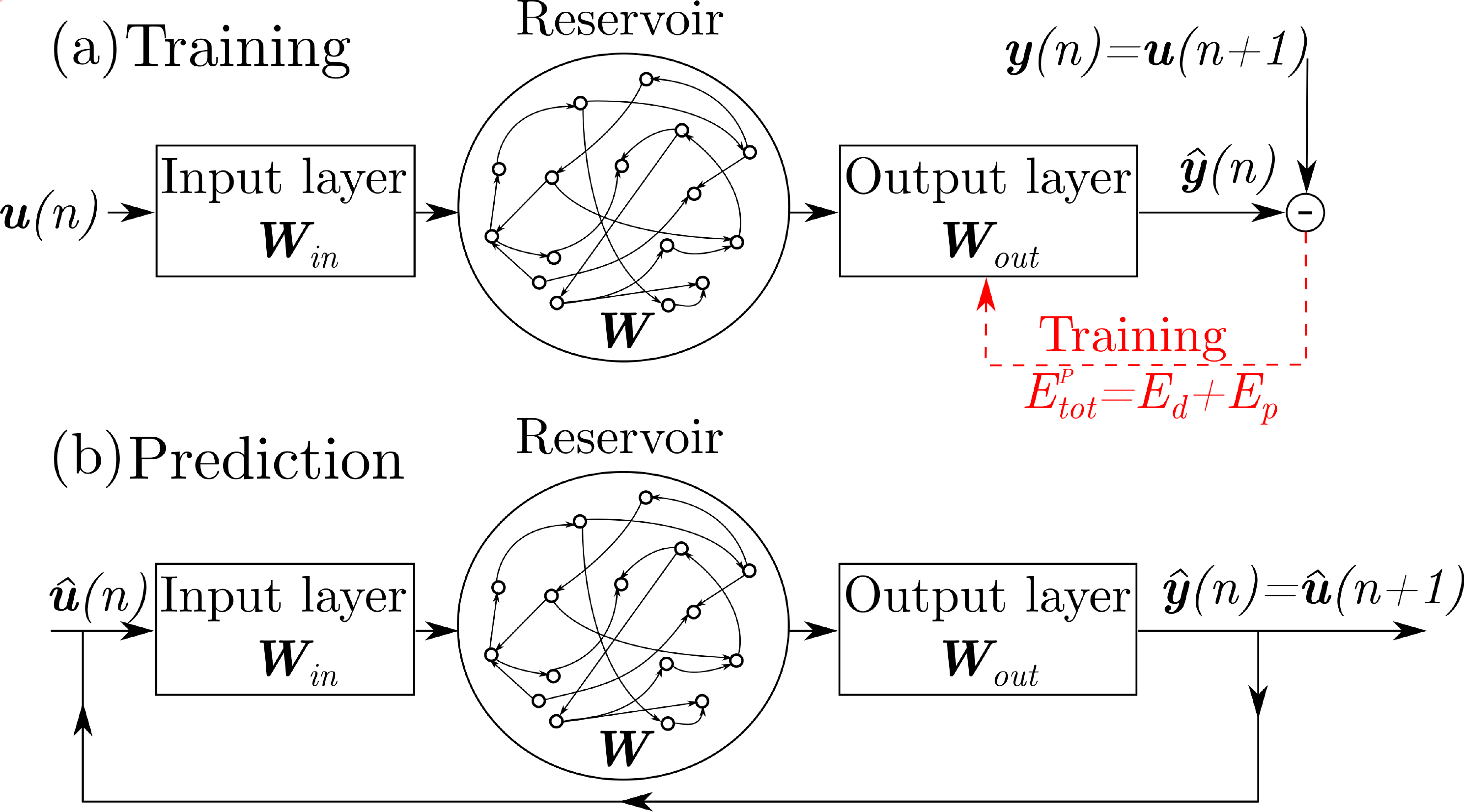}
    \caption{PI-ESN during (a) training and (b) prediction.}
    \label{fig:ESN_schema}
\end{figure}

To learn the reduced-order dynamics of shear turbulence, we constrain the physical knowledge of the governing equations into a reservoir computing data-driven method based on the Echo State Network~\citep{Jaeger2007,Lukosevicius2009} (ESN): The Physics-Informed Echo State Network~\citep{Doan2019} (PI-ESN). A schematic of the network is shown in Fig.~\ref{fig:ESN_schema}. 
We have training data with an input time series $\bm{u}(n)\in \mathbb{R}^{N_u}$ and a target time series $\bm{y}(n)\in \mathbb{R}^{N_y}$, where $n=0,1,2,\ldots, N_t$ are the discrete time instants that span from $0$ to $T=N_t\Delta t$. During prediction, the target at time $n$ becomes the input at time $n+1$, i.e., $\bm{u}(n+1)=\bm{y}(n)$. 
The training of the PI-ESN is achieved by (i) minimizing the error between the prediction, $\widehat{\bm{y}}(n)$, and the target data ${\bm{y}}(n)$ when the PI-ESN is excited with the input, $\bm{u}(n)$ (Fig. \ref{fig:ESN_schema}a), and (ii) enforcing that the prediction does not violate the physical constraints. To enforce (ii), we observe that 
a solution of the chaotic flow, $\bm{y}=\{a_i\}_1^9$, is such that the {\it physical error} (also known as the residual) is zero, i.e., $\mathcal{F}(\bm{y})\equiv \dot{\bm{y}}-\mathcal{N}(\bm{y})=0$.  %
To estimate the physical error beyond the training data, the PI-ESN is looped back to its input (Fig.~\ref{fig:ESN_schema}b) to obtain the predictions $\lbrace \widehat{\bm{y}} (n_p)  \rbrace_{p=1}^{N_p}$ in the time window with no training data, $(T+\Delta t) \leq t \leq (T+N_p\Delta t)$. The number of collocation points, $N_p$, is user-defined.  The physical error $\mathcal{F}({\widehat{\bm{y}}(n_p)})$ is evaluated to train the PI-ESN such that the sum of 
(i) the physical error between the prediction and the available data from $t=0$ to $t=T$, $E_d$, and 
(ii) the physical error for $t>T$, $E_p$, is minimized. Mathematically, we wish to find $\widehat{{\bm y}}(n)$ for $n=0,1,\ldots,N_t+N_p$ that minimizes 
\begin{equation}
E_{tot}^P =  \underbrace{\frac{1}{N_t} \sum_{n=1}^{N_t} \lvert\lvert\widehat{{\bm y}} (n) - {\bm y} (n)\lvert\lvert^2 }_{E_{d}} + \underbrace{  \frac{1}{N_p} \sum_{p=1}^{N_p} \lvert\lvert \mathcal{F}(\widehat{{\bm y}}(n_p))\lvert\lvert^2 }_{E_p}, 
\label{eq:EPhys}
\end{equation}
where $\lvert\lvert\cdot\lvert\lvert$ is the Euclidean norm. The PI-ESN is straightforward to implement because it requires only cheap residual calculations at the collocation points, i.e., it does not require the exact solution. 

\subsection{Network architecture}
The architecture of the PI-ESN follows that of the ESN, which consists of an input matrix $\bm{W}_{in}\in\mathbb{R}^{N_x \times N_u}$, which is a sparse matrix; a \textit{reservoir} that contains $N_x$ neurons that are connected by the  recurrent weight matrix $\bm{W}\in\mathbb{R}^{N_x \times N_x}$, which is another sparse matrix; and the output matrix $\bm{W}_{out}\in\mathbb{R}^{N_y\times N_x}$. The input time series, $\bm{u}(n)$, is connected to the reservoir through $\bm{W}_{in}$ to excite the states of the neurons, $\bm{x}$, as 
\begin{equation}
    \bm{x}(n+1) = \tanh \left( \bm{W} \bm{x}(n) + \bm{W}_{in} \bm{u}(n+1) \right)
\end{equation}
where $\tanh(\cdot)$ is the activation function. The output of the PI-ESN, $\widehat{\bm{y}}(n)$, is computed by linear combination of the reservoir states as $\widehat{\bm{y}}(n) = \bm{W}_{out} \bm{x}(n)$.

The matrices $\bm{W}_{in}$ and $\bm{W}$ are randomly generated and fixed~\citep{Lukosevicius2012}. Only $\bm{W}_{out}$ is trained to minimize \eqref{eq:EPhys}. Following~\cite{Pathak2018a}, each row of $\bm{W}_{in}$ has only one non-zero element, which is  drawn from a uniform distribution over $[-\sigma_{in},\sigma_{in}]$; $\bm{W}$ has an average connectivity $\langle d \rangle$, whose non-zero elements are drawn from a uniform distribution over the interval $[-1,1]$; and $\bm{W}$ is scaled such that its largest eigenvalue is $\Lambda\leq 1$, which ensures the Echo State Property \citep{Lukosevicius2012}.  
The training of the PI-ESN is achieved in two steps. First, the network is initialized by an output matrix, $\bm{W}_{out}$, that minimizes a data-only cost functional $E_{tot}^{NP} = E_d + \gamma || \bm{w}_{out,i} ||^2
\label{eq:err_data}$, where $\gamma$ is a Thikonov regularization factor and $\bm{w}_{out,i}$ denotes the $i$-th row of $\bm{W}_{out}$. This is the output matrix of the conventional ESN~\citep{Pathak2018}. 
Second, the physical error~\eqref{eq:EPhys} is minimized with the L-BFGS method~\citep{Byrd1995}, which is a quasi-Newton optimization algorithm. 

\section{Results}
\label{sec:results}
A grid search (supplementary material) provides the  hyperparameters $\Lambda = 0.9$, $\sigma_{in}=1.0$, $\langle d \rangle = 3$, $\gamma = 10^{-6}$, which enable accurate predictions in the range of $N_x=[500, 3000]$ neurons. Only $t=2500$ time units (equivalent to $t^+\approx 61$) in the window $t=[11500, 14000]$ (equivalent to $t^+\approx[280, 341]$ in the grey box of Fig.~\ref{fig:MFE_evol}) are used for training. The data beyond this time window is used for validation only. We use $N_p=5000$ collocation points (equivalent to $t=1250$ or $t^+\approx30.5$), which provide  a sufficient number of predictions beyond the training data  with a relatively low computational time. %

\subsection{Long-term statistical behaviour}
\label{sec:res_LT}
\begin{figure}[!ht]
    \centering
    \includegraphics[width=0.7\textwidth]{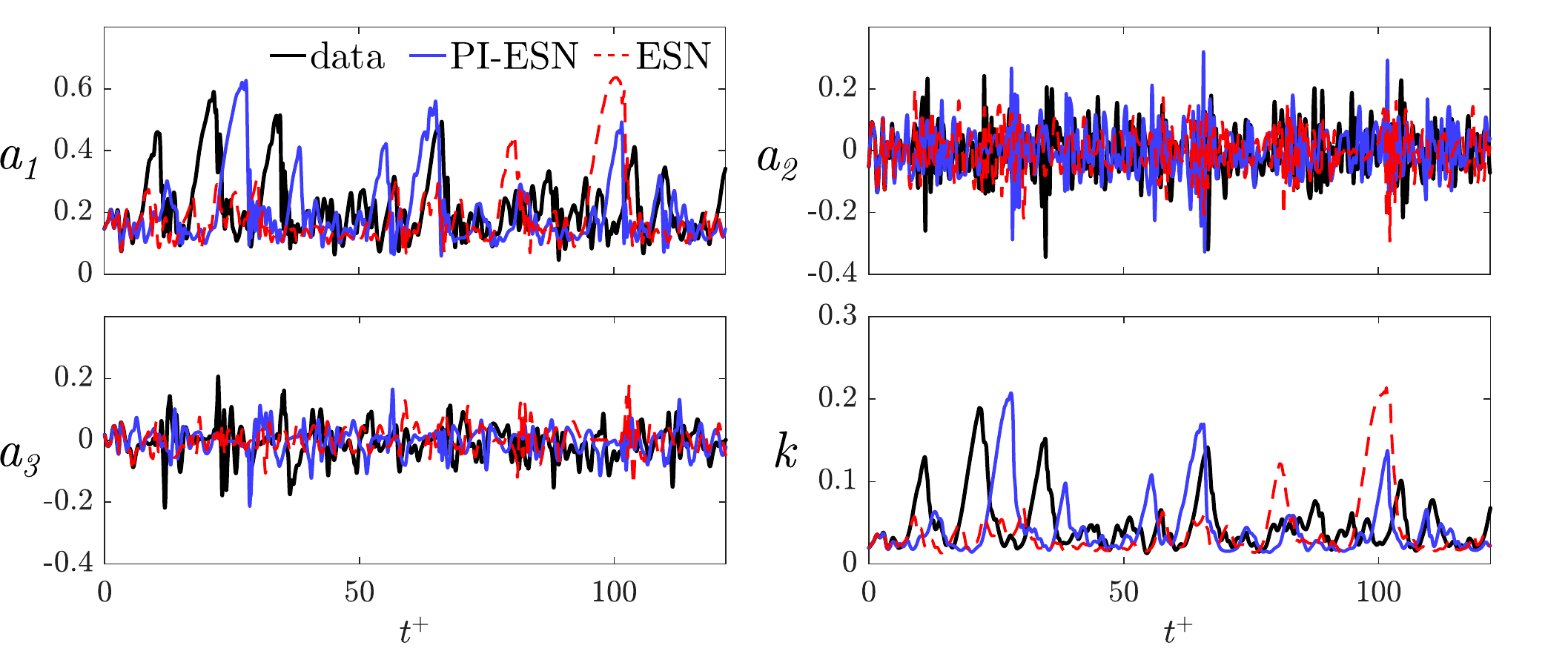}
    \caption{Long-term evolution of the modes' amplitudes: exact evolution (thick black lines), ESN solution (dashed red lines), PI-ESN solution (blue full lines) with reservoir of 2500 units.}
    \label{fig:MFE_LT_pred}
\end{figure}

A long-term prediction of the modes' amplitudes by the trained ESN and PI-ESN is shown in Fig. \ref{fig:MFE_LT_pred} for a reservoir of 2500 units. This prediction is made for 5000 time units, corresponding to approximately 120 Lyapunov times. It is seen that overall, both the ESN and the PI-ESN qualitatively reproduce a dynamics similar to the original MFE system. In particular, both the ESN and PI-ESN exhibit transition towards a quasi-laminar state with large growth of $a_1$ and the associated increase in $k$.

\begin{figure}[!ht]
    \centering
    \includegraphics[width=0.75\textwidth]{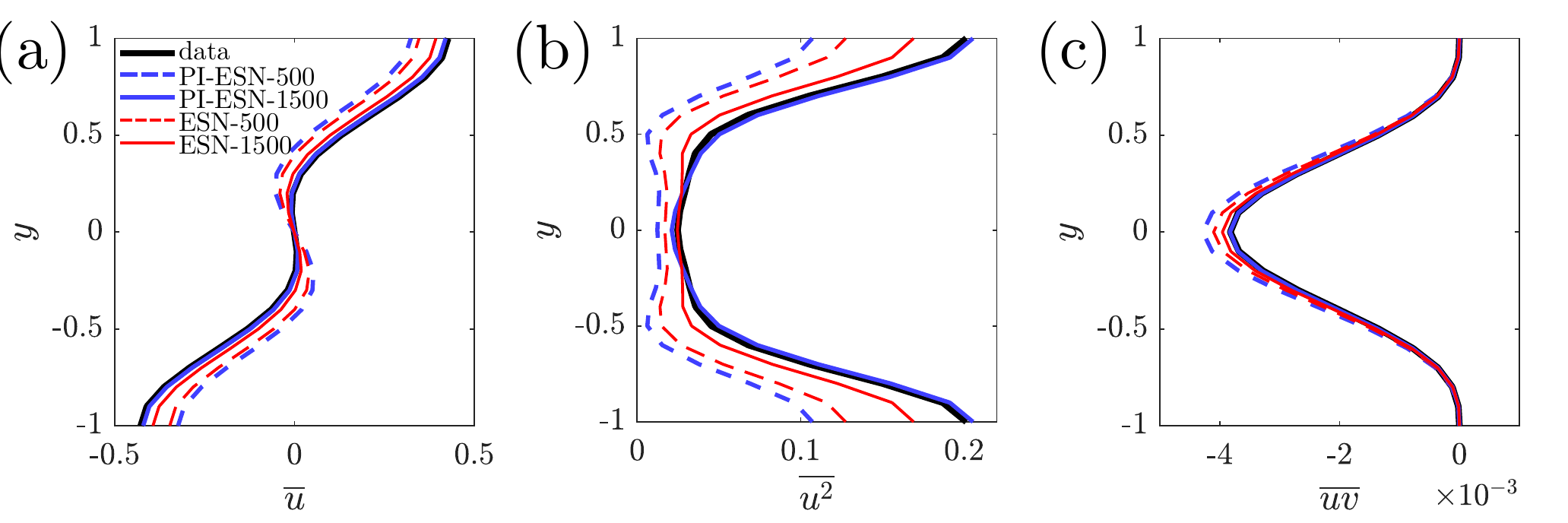}
    \caption{Profile of (a) mean streamwise velocity, (b) variance and (c) mean Reynolds stress.}
    \label{fig:MFE_LT_velo_stats}
\end{figure}

To make a quantitative comparison, the statistics of the velocity are shown in Fig. \ref{fig:MFE_LT_velo_stats} for ESNs and PI-ESNs of different reservoir sizes. The statistics, which are collected for 5000 time units, are computed by averaging the velocity along the two periodic directions, $x$ and $z$, and time.

Figure \ref{fig:MFE_LT_velo_stats} shows that for small reservoirs (500 units), both the ESN and PI-ESNs do not accurately reproduce the mean streamwise velocity profile and its variance. This is because there are not enough neurons in the reservoir to correctly reproduce the intricate dynamics of the MFE model and its long-term evolution. However, as the reservoir size is increased to 1500 units, the PI-ESN accurately captures the mean velocity profile, its variance and the Reynolds stress. On the other hand, the data-only ESN solution is less accurate. A larger reservoir of 2500 units is required for the data-only ESN to accurately reproduce the statistics of the MFE model (result not shown). This highlights how the physical constraint augments the information used during the training of the PI-ESN allowing it to better capture the long-term evolution compared to a data-only ESN.
The PI-ESN approach needs a relatively small training dataset of 2500 time units to  learn the long-term statistics of the system. Compared to previous studies with Long Short-Term Memory units, this training data is two to three orders of magnitude smaller \cite{Srinivasan2019}.

\subsection{Short-term extreme events prediction}
\label{sec:res_EE}
\begin{figure}[!ht]
    \centering
    \includegraphics[width=0.70\textwidth]{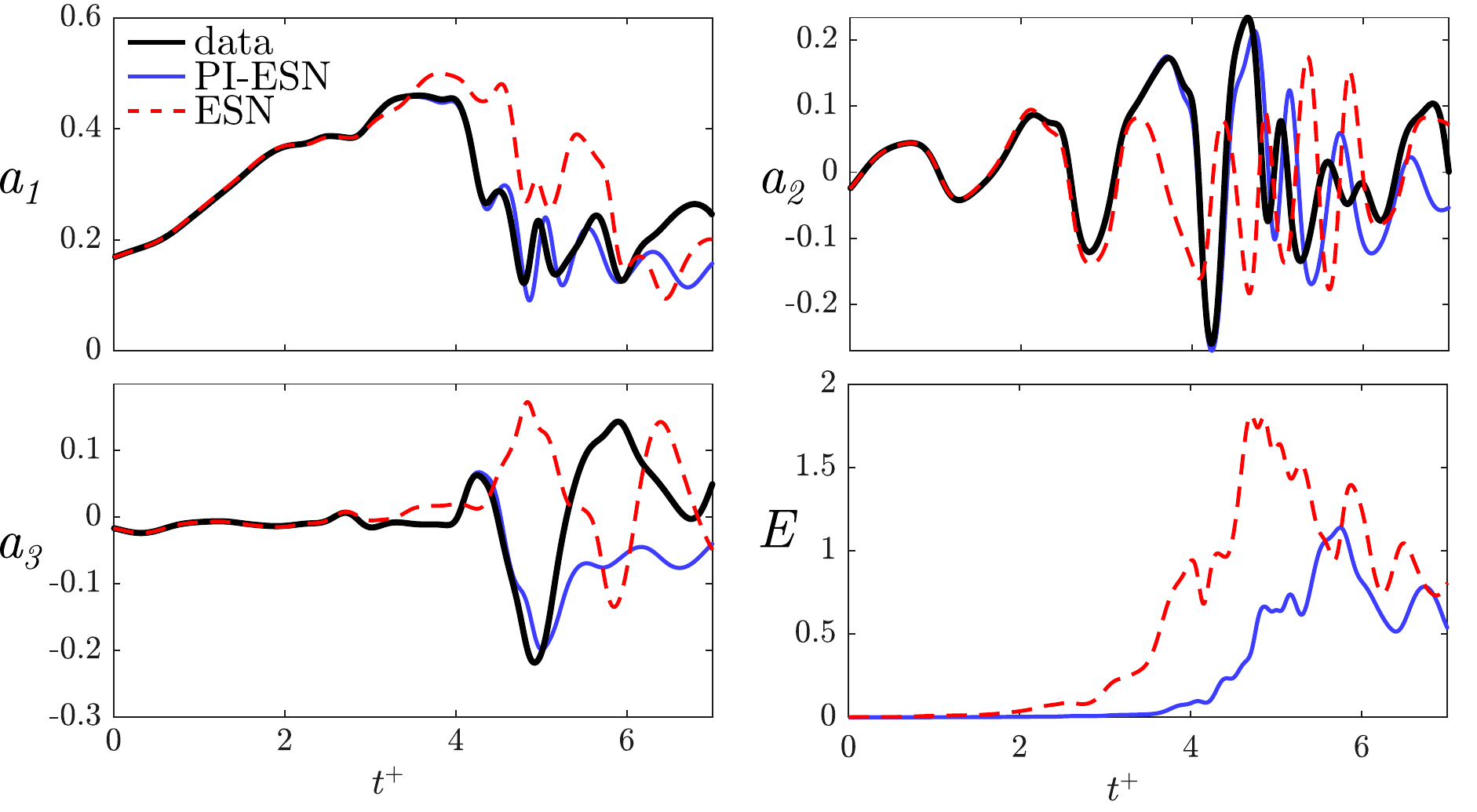}
    \caption{Evolution of  modes $a_1$, $a_2$, $a_3$ during the extreme event of Fig. \ref{fig:MFE_evol}: exact evolution in solid black line; prediction with physics-informed echo state network (PI-ESN) in solid blue line; and prediction with fully data-driven echo state network (ESN) in dashed red line. The reservoir has $N_x=3000$ neurons. $E$ is the error (Eq. \eqref{eq:normError}).}
    \label{fig:MFE_ESN_comp}
\end{figure}

\begin{figure}[!ht]
    \centering
    \includegraphics[width=0.75\textwidth]{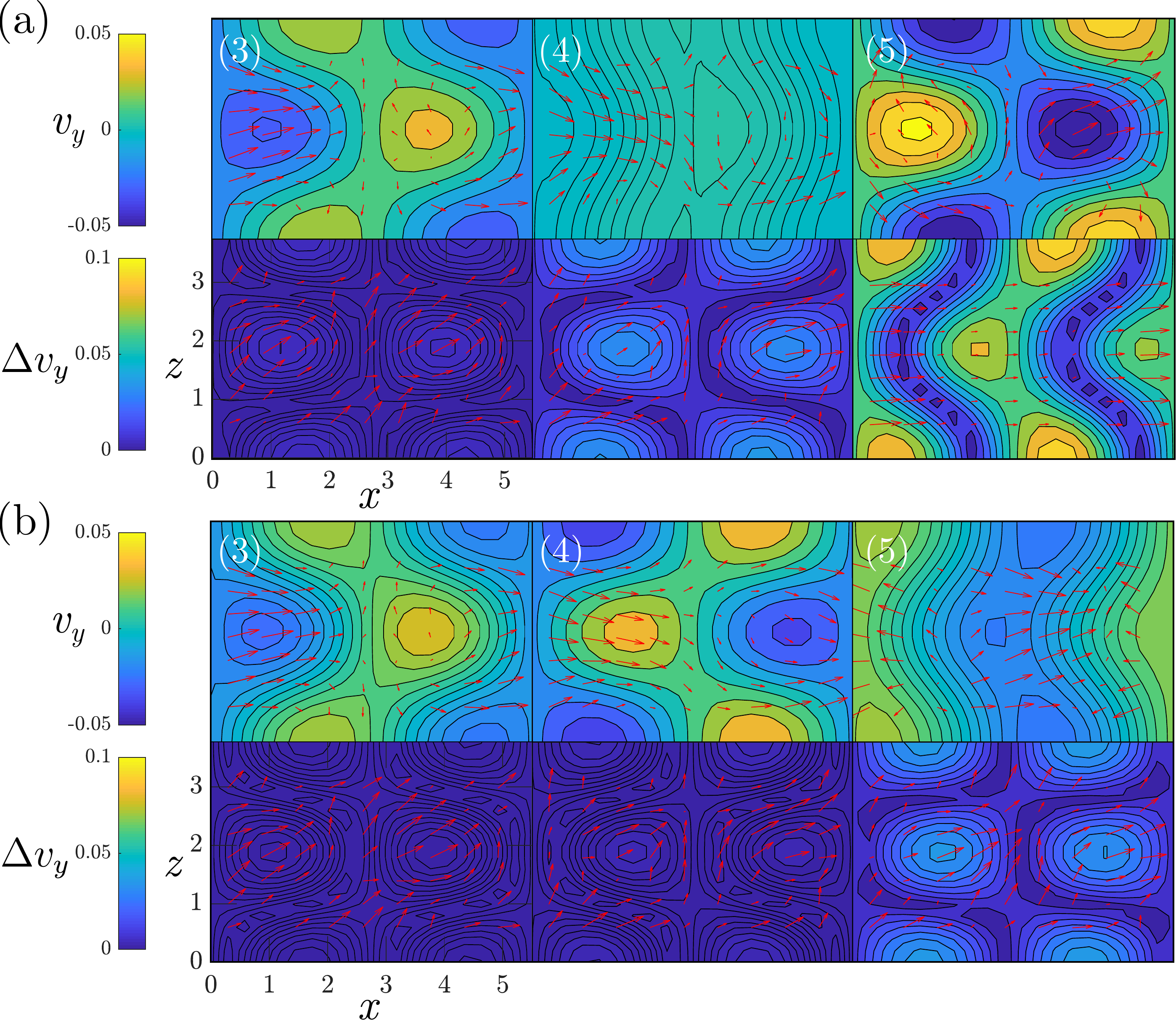}
    \caption{Evolution of the velocity field (top rows) and the normalized error on the out-of-plane velocity (bottom rows) in the velocity field in the mid-$y$ plane at the same time instants as panels (3)-(5) of Fig. \ref{fig:MFE_ESN_comp}.  Predictions from (a) the conventional ESN and (b) the PI-ESN. The arrows indicate the in-plane velocity ($x$-$z$ directions) and the coloured contour indicates the out-of-plane velocity. The panels correspond to $t^+ = t\lambda_{\max} \approx 2.14,~3.21,~4.27$ in Fig. \ref{fig:MFE_ESN_comp}. 
    }
    \label{fig:MFE_ESN_comp_2D}
\end{figure}

Figure~\ref{fig:MFE_ESN_comp} shows the evolution of three  modes  during the extreme event in the dashed red box of the top panel of  Fig.~\ref{fig:MFE_evol}. The PI-ESN solution (solid blue line) and the conventional ESN solution (dashed red line) are computed with a reservoir of $N_x=3000$ units. Both solutions are compared against the exact solution from numerical integration (solid black line). The normalized error  between the exact evolution and the PI-ESN/ESN predictions is computed as
\begin{equation}
\label{eq:normError}
    E(n) = \frac{||\bm{y}(n) - \widehat{\bm{y}}(n) ||}{\sqrt{\frac{1}{N_t}\sum_{n=1}^{N_t} || \bm{y}(n)|| } } 
\end{equation} where the denominator of the error cannot be zero because the fixed point of the MFE model has $||\bm{y}|| = 1$ and the system has unsteady chaotic oscillations.
Although the same training data is used for both the PI-ESN and the conventional ESN, the PI-ESN has a significantly higher extrapolation-in-time capability than the conventional ESN. To compare the performance, we define the predictability horizon as the time required for $E$ to exceed the threshold $0.2$ from the same initial condition. The predictability horizon of the PI-ESN is $\approx 2$ Lyapunov times longer than the predictability horizon of the conventional ESN. This improvement is achieved by enforcing the prior physical knowledge of the  flow, whose evolution must fulfil the momentum and mass conservation laws.  
As shown in Fig.~\ref{fig:MFE_ESN_comp}, until $t^+\approx 2.14$, both ESN and PI-ESN accurately predict the flow evolution. The predicted solution from the conventional ESN starts  diverging from the exact evolution at $t^+ \approx 3.21$, which leads to a completely different solution during the extreme event. On the other hand, the PI-ESN is able to time-accurately predict the occurrence and the amplitude of the extreme event. After the event has occurred, the solution diverges because the butterfly effect is significant. 
The velocity fields predicted by the conventional ESN and PI-ESN are shown in Fig.~\ref{fig:MFE_ESN_comp_2D}a,b, respectively. The solutions are shown at the same times as the exact solution in panels (3)-(5) of Fig. \ref{fig:MFE_evol}.  The bottom rows of Fig. \ref{fig:MFE_ESN_comp_2D}a,b show the normalized absolute error between the predicted velocity field and the exact velocity field. 
The discrepancy in the  velocity field is mainly due to the error on the prediction of the downstream vortex mode, $a_3$ (Fig. \ref{fig:MFE_ESN_comp}). On one hand, because no physical knowledge is constrained in the conventional ESN, the sign and amplitude of  $a_3$ are incorrectly predicted, which means that the out-of-plane velocity evolves in the opposite direction of the exact solution. On the other hand, the PI-ESN is able to predict satisfactorily  both the in-plane velocity and the out-of-plane velocity during the extreme event.

\begin{figure}[!ht]
    \centering
    \includegraphics[width=0.5\textwidth]{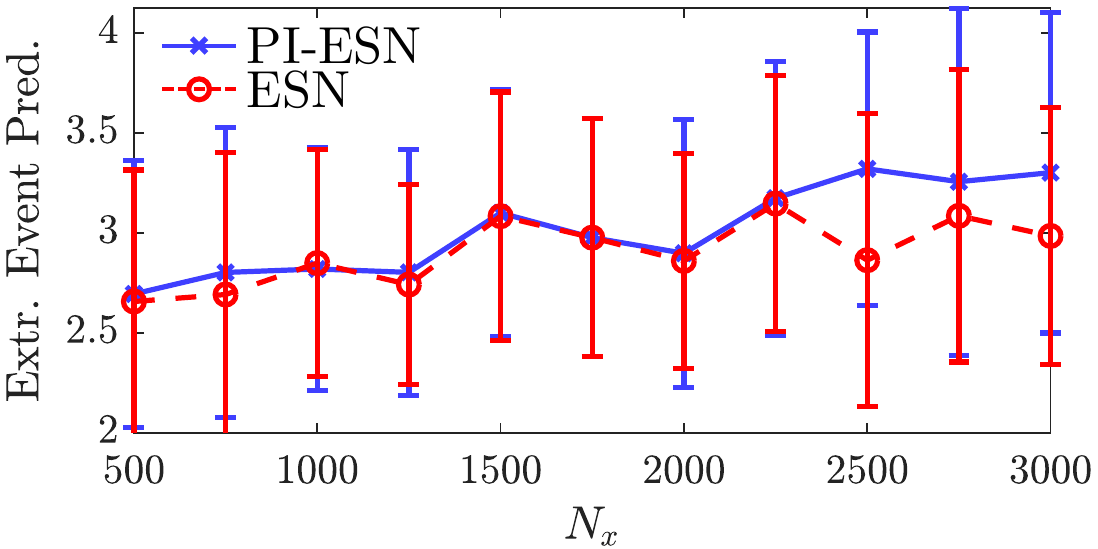}
    \caption{Comparison of the average predictability horizons of the PI-ESN (full blue line with crosses) and ESN (dashed red line with circles) for all the extreme events in the dataset. The errorbars indicate half a standard deviation. $N_x$ is the number of neurons.}
    \label{fig:MFE_comp_predRE}
\end{figure}

To quantitatively assess the robustness of these results, we compute the average predictability horizon of the machines with no further training.  We follow the following steps: 
(i) by inspection of Fig.~\ref{fig:MFE_evol}, we define events as extreme when their kinetic energy is $k\geq0.1$; 
(ii) we identify the times when all the extreme events start in the dataset of Fig.~\ref{fig:MFE_evol}; 
(iii) for each time, the exact initial condition at $t^+ \approx 0.61$ just before the time instant in which the extreme events starts is inputted in the PI-ESN and ESN; 
(iv) the machines are time evolved to provide the prediction; and 
(v) the predictability time is computed by averaging over all the extreme events in the dataset. The mean predictability time and the standard deviation are computed with validation data containing twenty extreme events. The results are parameterized with the size of the reservoir, $N_x$ (Fig. \ref{fig:MFE_comp_predRE}).
On one hand, for small reservoirs ($N_x\lesssim2000$), the performances of the ESN and PI-ESN are comparable. This means that the performance is more sensitive to the data cost functional, $E_d$, than the physical error, $E_p$.
On the other hand, for larger reservoirs ($N_x\gtrsim2000$), the physical knowledge is fully exploited by the PI-ESN. This means that the performance becomes markedly sensitive to the physical error, $E_p$. This results in an improvement in the average predictability of $\approx 1.5$ Lyapunov times. Because an extreme event takes $\approx 3$ Lyapunov time on average, the improved predictability time of the PI-ESN is key to the time-accurate prediction of the occurrence and amplitude of the extreme events. 

\subsection{Robustness to noisy data}
\label{sec:res_noise}
Because the accuracy of the ESN and PI-ESN---and more generally, of any data-driven method---depends on quality of the training data, the effect of additive noise is analysed. Noise is added to the training dataset presented in Fig. \ref{fig:MFE_evol} with Signal to Noise Ratios (SNR) of 30 and 40dB, which are representative of experimental noise~\cite{Ouellette2006}. The ESN and PI-ESN are trained on the noisy data, and the accuracy of the long- and short-term prediction is analysed. The dataset length and hyperparameters are the same as those of Secs.~\ref{sec:res_LT} and \ref{sec:res_EE}.

\begin{figure}[!ht]
    \centering
    \includegraphics[width=0.7\textwidth]{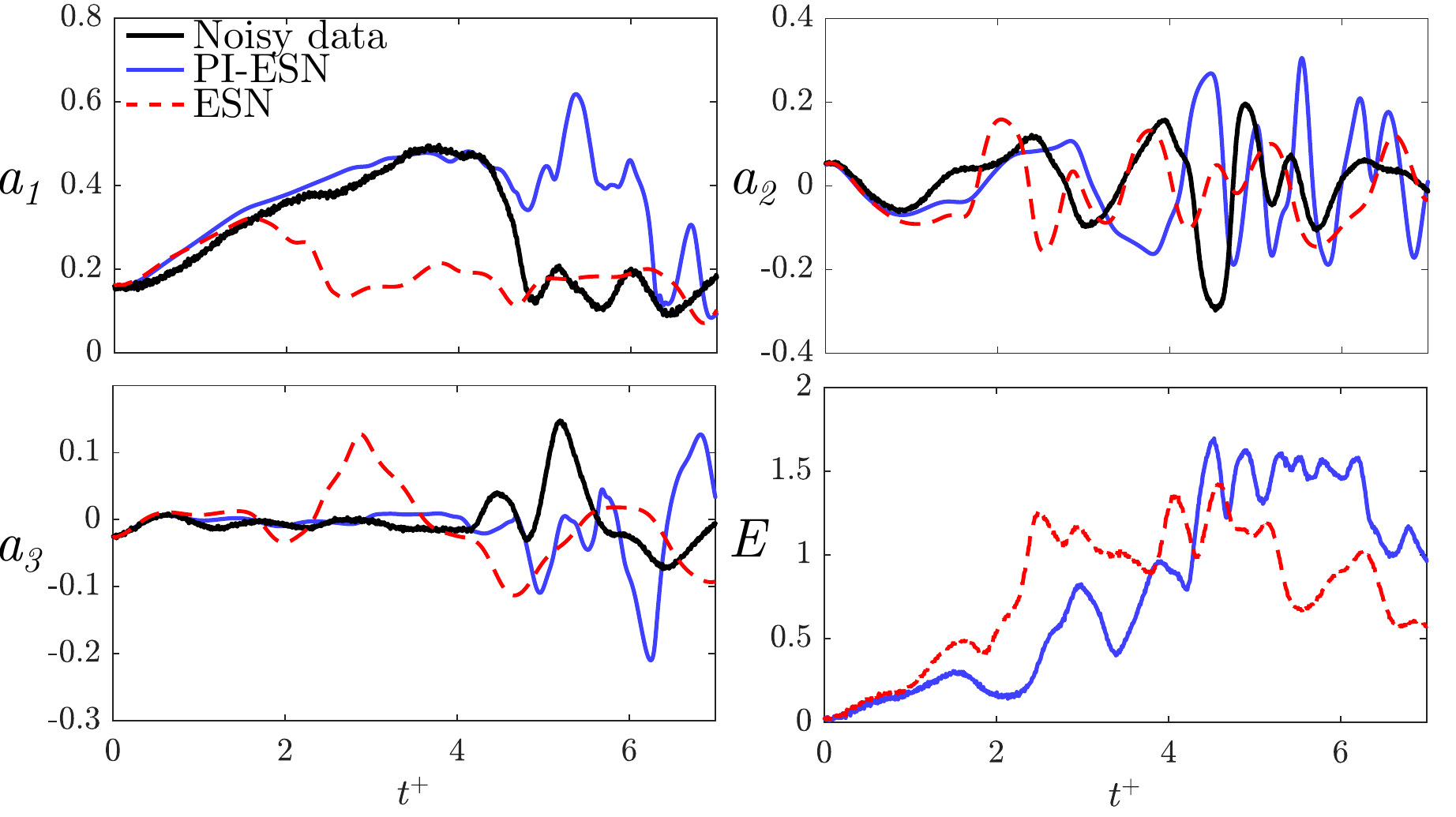}
    \caption{Evolution of $a_1$, $a_2$, $a_3$ during the extreme event of Fig. \ref{fig:MFE_evol}: noisy evolution in solid black line; prediction with physics-informed echo state network (PI-ESN) in solid blue line; and prediction with fully data-driven echo state network (ESN) in dashed red line. The reservoir has $N_x=3000$ neurons. $E$ is the error (Eq. \eqref{eq:normError}). The noisy data corresponds to the case with SNR=30.}
    \label{fig:MFE_ESN_comp_noise}
\end{figure}

Figure~\ref{fig:MFE_ESN_comp_noise} shows the evolution of three modes, the normalized error and the noisy solution. Similarly to the case of zero noise (Fig. \ref{fig:MFE_ESN_comp}), the PI-ESN is quantitatively more predictive than the data-only ESN. For example, the PI-ESN predicts the increase of the $a_1$ mode, which is physically the quasi-laminar profile forming during the extreme event. Although after 4 Lyapunov times the PI-ESN diverges due to the high sensitivity of chaotic flows to noise, the overall statistics are well predicted, as discussed next. One possible strategy to improve the performance is to increase the number of collocation points in the training of the PI-ESN.  

\begin{figure}[!ht]
    \centering
    \includegraphics[width=0.5\textwidth]{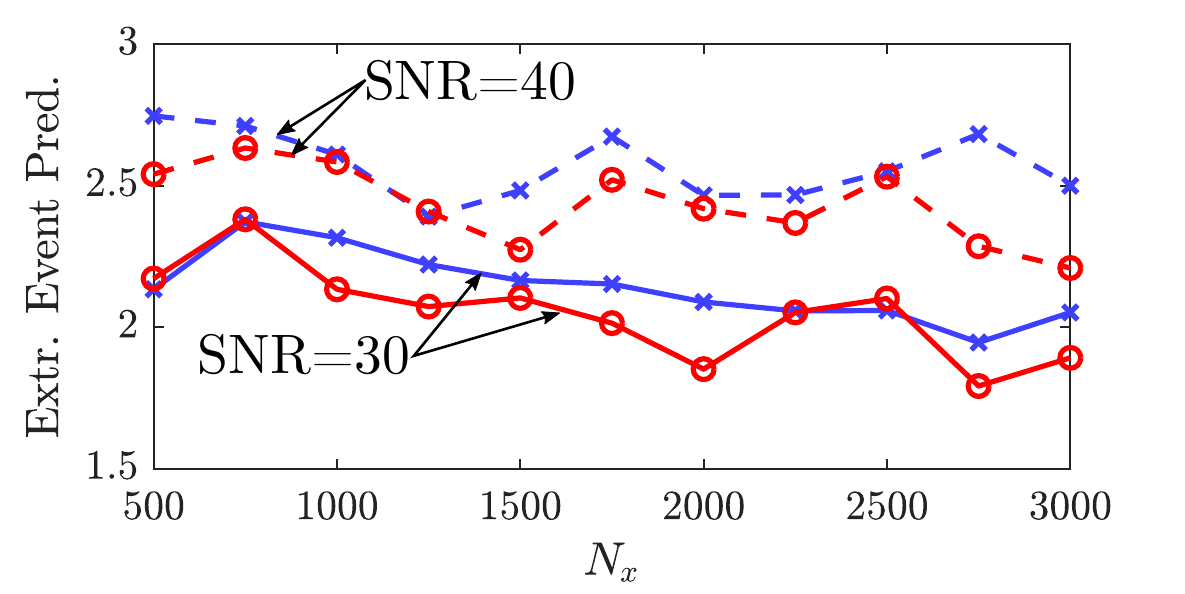}
    \caption{Comparison of the average predictability horizons of the PI-ESN (blue full and dashed lines with crosses) and ESN (red full and dashed lines with circles) for all the extreme events in the dataset when trained with noisy database. Full lines: SNR=40, dashed lines: SNR=30. $N_x$ is the number of neurons.}
    \label{fig:MFE_comp_noise_predRE}
\end{figure}

The statistical assessment is shown in Fig. \ref{fig:MFE_comp_noise_predRE}. Compared to the noise-free case of Sec. \ref{sec:res_EE}, the predictability time of extreme events becomes smaller as the noise level becomes larger. This loss in accuracy is due to the noise in the training data. Both the ESN and PI-ESN, minimize the error on the noisy data, $E_d$, which causes the machines to try to fit the noisy data, resulting in a loss in predictive accuracy. This loss becomes more significant as the noise increases. If the number of neurons in the reservoir is further increased, the ESN and, to a less extent, the PI-ESN will start overfitting the noise. The PI-ESN mitigates the loss of accuracy through the the physical loss, $E_p$, which acts as a regularization term that filters out the noise. This  allows the PI-ESN to maintain a higher accuracy in the prediction of the extreme event (Fig. \ref{fig:MFE_ESN_comp}). As a result, the PI-ESN provides a longer accurate prediction of the extreme events, by up to $0.4$ Lyapunov time, compared to the data-only ESN. This effect bercomes more significant as the number of neurons in the reservoir increases. In other words, the physical constraint in the PI-ESN acts as a denoiser. This overcomes the lack of robustness of the data-only ESN, which cannot discriminate the noise from the actual flow dynamics.

\begin{figure}[!ht]
    \centering
    \includegraphics[width=0.75\textwidth]{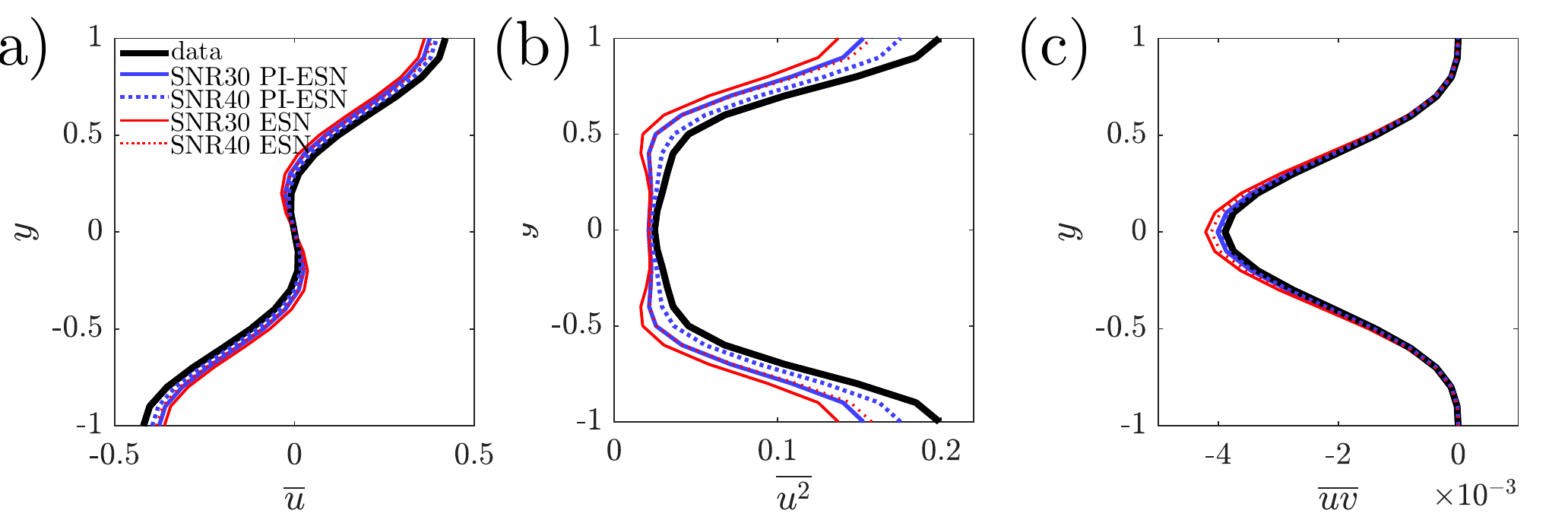}
    \caption{Profile of (a) mean streamwise velocity, (b) variance and (c) mean Reynolds stress.}
    \label{fig:MFE_noiseLT_velo_stats}
\end{figure}

Finally, Fig.~\ref{fig:MFE_noiseLT_velo_stats} compares the statistics of the velocity obtained from a long-term prediction of 5000 time units for a reservoir size of 1500 units (similarly to the noise-free case of Fig. \ref{fig:MFE_LT_velo_stats}). Consistently with the short-term performance, the accuracy on the statistics decreases as the noise in the training data. For both noise levels, the PI-ESN outperforms the data-only ESN.

\section{Final discussion and future directions}
\label{sec:conclusion}
We propose a physics-informed echo state network (PI-ESN), which combines empirical modelling, based on reservoir computing, with physical modelling, based on conservation laws, to time-accurate predict extreme events in a chaotic flow and its statistical behaviour. 
We  compare the performance of the PI-ESN with a fully-data driven echo state network (ESN). 
The former is a physics-constrained machine, whereas the latter is a physics-unaware machine because it is trained with data only. 
In the PI-ESN, the physical error from the conservation laws is minimized beyond the training data.  
This brings in crucial information, which is exploited in two ways:
(i) with the same amount of available data, the PI-ESN solution is accurate for a longer time than the conventional ESN solution, i.e., the information from physical knowledge enhances {\it extrapolation}; and %
(ii) less data is required to obtain the same accuracy as the conventional ESN. 
 Here, we take advantage of property (i) for the prediction of extreme events in a model of turbulence and the statistics of velocity and Reynolds stress. 
Extreme events may be generally rare, which makes it difficult for physics-unaware data-driven methods to be trained. 
On the other hand, constraining the physics enables the PI-ESN to predict extreme events that cannot be inferred from the data only. Additionally, we show that long-term statistics are also better captured while the PI-ESN is trained with a relatively short time series, which is attractive when only small datasets are available. 
Finally, the approach also shows robustness with respect to noise.
Physics-constrained reservoir compuring is being extended to higher dimensional systems.
This study opens up new possibilities for synergistically enhancing  data-driven methods with physical principles for the time-accurate prediction of extreme events in chaotic flows. 

\section*{Declaration of Interests}
 The authors report no conflict of interest. 

\section*{Acknowledgements}
The authors acknowledge the support of the TUM Institute for Advanced Study funded by the German Excellence Initiative and the EU 7$^{th}$ Framework Programme n. 291763. L.M.  acknowledges support from the Royal Academy of Engineering Research Fellowships Scheme.

\bibliographystyle{apsrev4-2}
%

\end{document}